# Open-stub based spurious harmonic suppression method for microstrip coupled line filter

Mohammad Hossein Koohi Ghamsari[1] - Erfan Azizkhani[2]- Javad Ghalibafan[3]

[1] Department of Electrical Engineering, Sharif University of Technology, Tehran, Iran
[2] Department of Electrical Engineering, Sharif University of Technology, Tehran, Iran
[3] Faculty of Electrical Engineering and Robotic, Shahrood University of Technology, Shahrood, Iran

**ABSTRACT**

In this paper, first, we review a straightforward analytical technique based on image impedance concept for designing traditional microwave microstrip coupled line filters using distributed elements. In the introduced approach, we characterize a quarter-wave coupled line section, and then these discrete sections can be connected in series to synthesis the final desired frequency response. Next, we use a novel open-stub based technique to suppress spurious harmonic frequencies. Finally, using proposed technique, we design and simulate a band pass filter (BPF). The simulation results prove the usefulness of the proposed technique.

## 1. INTRODUCTION

Parallel-coupled microstrip bandpass filters are widely used in RF front ends of wireless communication systems due to its planar structure, synthesis procedure, low cost, tightened capacitive coupling, easy integration and wide range of fractional bandwidth [1]. Design of a compact band pass filter using parallel-coupled half- wavelength microstrip resonators with improved stopband skirt characteristics and harmonic suppression is presented in [2]. [3] introduces a novel folded T-shaped open stub for harmonic suppression in compact size GaAs MMIC filters.

In this paper, we report a simple and efficient method for spurious harmonic suppression in parallel coupled line filters. The proposed method can also be used for different characteristics and topologies of microstrip filters.

Section 2 describes the principles and design procedure of the microstrip coupled line filter and analytical approach for analysis the effect of including open stubs in the structure. Section 3 shows an example of the proposed BPF. The conclusion is in Section 4.

## 2. ANALYSIS AND DESIGN

In filter analysis and design literature there is a number of well-known techniques for synthesis of different characteristics of frequency response. Here we use the method based on image impedance concept. Figure 1 shows a simple equivalent circuit of a basic unite of a coupled line.

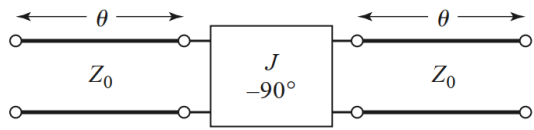

**Fig. 1.** Equivalent circuit of a coupled line section

Calculating ABCD parameters of the equivalent circuit and admittance inverter by considering it as a quarter-wave length of transmission of characteristic impedance, $1/J$, and using image impedance concept, after some mathematical calculations we get [4]

$$Z_0 J_1 = \sqrt{\frac{\pi \Delta}{2 g_1}} \qquad (1)$$

$$Z_0 J_n = \frac{\pi \Delta}{2\sqrt{g_{n-1} g_n}} \quad \text{for } n = 2, 3, \ldots, N \qquad (2)$$

$$Z_0 J_{N+1} = \sqrt{\frac{\pi\Delta}{2g_N g_{N+1}}} \qquad (3)$$

Which are design equations for a bandpass filter with N+1 coupled line sections. In these relations, $g_n$ is low-pass prototype value and $J_n$ is admittance inverter constant. The even- and odd-mode characteristic impedances for each section are found from

$$Z_{0e} = Z_0[1 + JZ_0 + (JZ_0)^2] \qquad (4)$$
$$Z_{0o} = Z_0[1 - JZ_0 + (JZ_0)^2] \qquad (5)$$

Attenuation versus normalized frequency for equal-ripple filter prototypes with 0.5 dB ripple level or other levels can be found from experimental graphs.

From [3], one optimum approach of using open-stubs for this application is as figure 2 which yields a T-shaped coupled line unit.

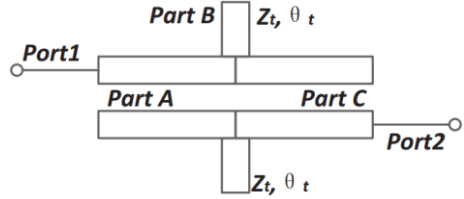

**Fig. 2.** Using open-stubs in a T-shaped configuration for a coupled line unit

The open stubs in part B could provide a transmission zero at higher frequency. Figure 2 can be modeled as combination of two parallel-coupled lines ($Z_c, \theta_c = 90°$), two open stubs ($Z_t, \theta_t$), and a J admittance inverter (1/J, 90°) as figure 3.

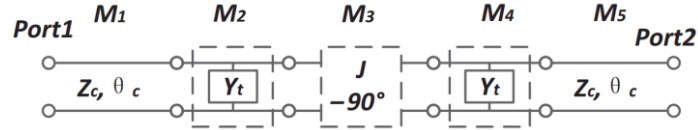

**Fig. 3.** Equivalent circuit of a T-shaped coupled line section

The ABCD parameters of this equivalent circuit can be computed using ABCD matrices of each section as follows

$$M_1 = M_5 = \begin{bmatrix} \cos(\theta_c) & iZ_c \sin(\theta_c) \\ iY_c \sin(\theta_c) & \cos(\theta_c) \end{bmatrix} \qquad (6-1)$$

$$M_2 = M_4 = \begin{bmatrix} 1 & 0 \\ iY_t \tan(\theta_t) & 1 \end{bmatrix} \qquad (6-2)$$

$$M_3 = \begin{bmatrix} 0 & -i/J \\ -i/J & 0 \end{bmatrix} \qquad (6-3)$$

$$M_{TCouple} = M_1 M_2 M_3 M_4 M_5 \qquad (6-4)$$

$$M_{TCouple} = \begin{bmatrix} -\dfrac{\sin(\theta_c)}{JZ_t} & i\left(1 - \dfrac{\sin^2(\theta_c)}{J^2 Z_t^2}\right) JZ_0^2 \\ i\dfrac{1}{JZ_c^2} & -\dfrac{\sin(\theta_t)}{JZ_t} \end{bmatrix}$$

## 3. SIMULATION AND RESULTS

For designing a BPF with desired frequency response characteristics, one can start with a single quarter-wave coupled line section and choose transfer function prototype (Butterworth, Chebyshev, etc.) to get the initial basic parameters of the J admittance and by connecting the proposed units in series, we can implement the desired BPF. Following the design procedure as discussed in section 2, we designed a BPF with center frequency of 2 GHz as shown in figure 4.

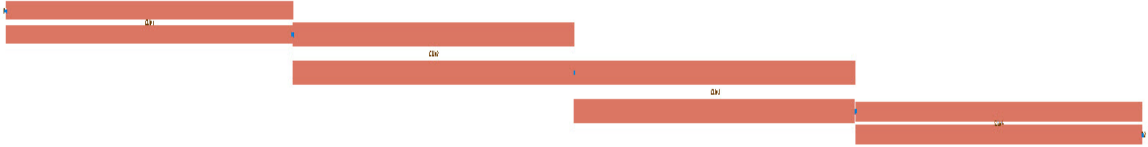

**Fig. 4.** Traditional microstrip parallel coupled line filter

The suppression of spurious harmonic response is obtained by inclusion of T-shaped open-stubs to the implemented filter. Because there are many unknown parameters, we need to optimize the proposed filter layout by full-wave electromagnetic simulation software. The final result is shown in figure 5.

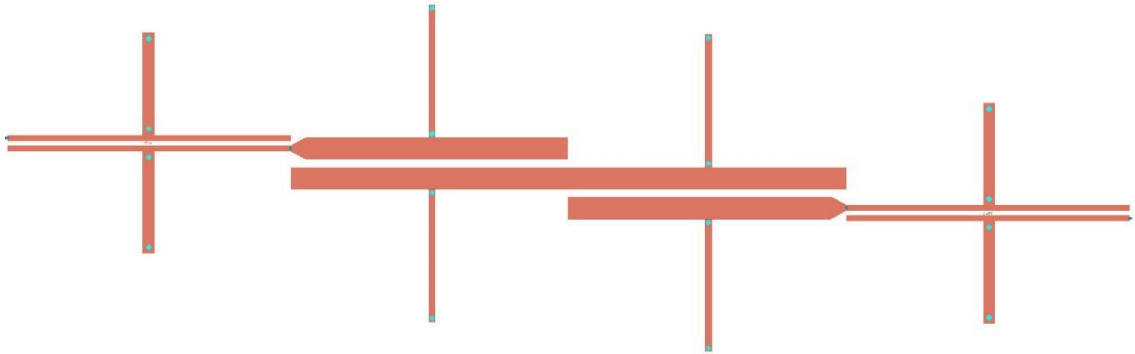

**Fig. 5.** Optimum configuration of coupled line filter with T-shaped open-stubs after applying optimizations and mitered bends technique

Position and length of open-stubs can be optimized for best return loss and insertion loss. For the requirement of compact implementation, such as in MMICs, we can use optimum folded T-shaped open-stubs, as shown in figure 6. Furthermore, it can be shown that by careful manipulation of open-stubs, suppression of multiple spurious harmonic frequencies is also possible.

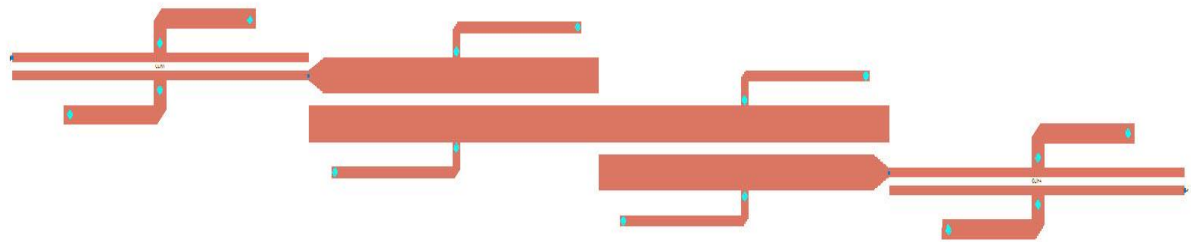

**Fig. 6.** Coupled line filter with folded T-shaped open-stubs

The simulated S-parameters of the traditional and proposed filter are presented in figure 7 and figure 8. As it is shown in the proposed filter response, the return loss and insertion loss is increased about 19 dB and 17 dB in the second and third harmonic pass bands, respectively.

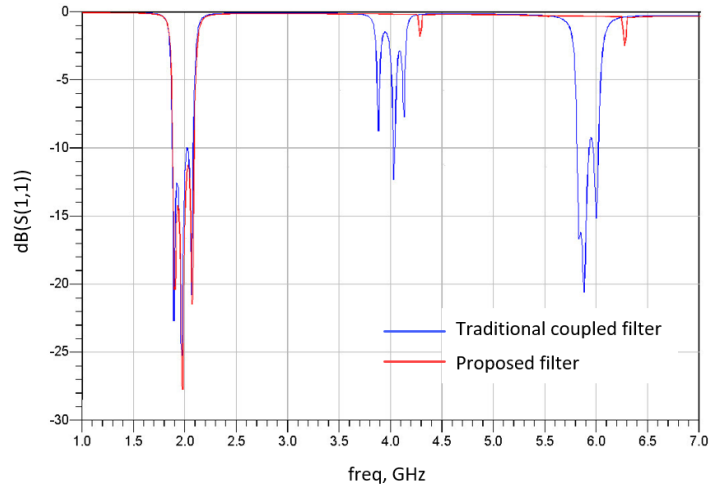

**Fig. 7.** S$_{11}$ of traditional and proposed coupled line filter

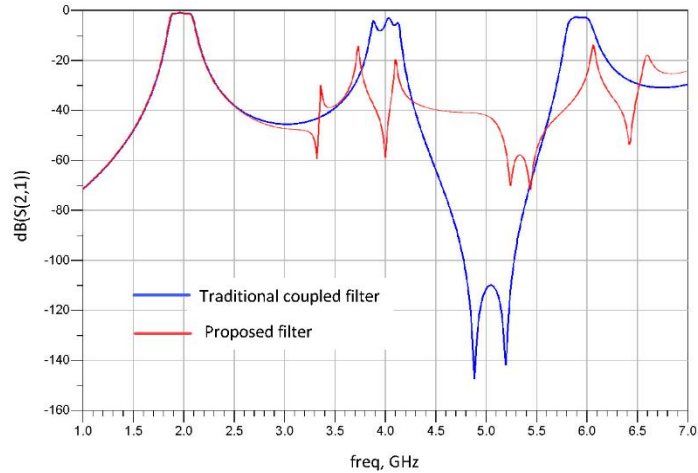

**Fig. 8.** S$_{21}$ of traditional and proposed coupled line filter

**4. CONCLUSION**

We reviewed a simple design approach for printed parallel-coupled filter which is popular for it's compactness and good frequency response in terms of low insertion loss and wide stopband characteristics. Next, we used T-shaped open-stubs for suppression of spurious response at harmonic frequencies, which could cause interference with other electronic components. The proposed method does not require any significant change in the designed filter with traditional design procedure. The simulation results show advantages of proposed filter structure over traditional configuration.